\documentclass{elsarticle}
\biboptions{sort&compress}

\usepackage{hyperref}

\begin{document}

\begin{frontmatter}

\title{First-principles calculations of 
the magnetocrystalline anisotropy of the prototype 2:17 cell boundary phase Y(Co$_{1-x-y}$Fe$_x$Cu$_y$)$_5$}

\author[mymainaddress]{Christopher E.\ Patrick\corref{mycorrespondingauthor}}
\cortext[mycorrespondingauthor]{Corresponding author}
\ead{c.patrick.1@warwick.ac.uk}

\author[mysecondaryaddress]{Munehisa Matsumoto}

\author[mymainaddress]{Julie B.\ Staunton}

\address[mymainaddress]{Department of Physics, University of Warwick, Coventry CV4 7AL, United Kingdom}
\address[mysecondaryaddress]{Institute for Solid State Physics, University of Tokyo, Kashiwa 277-8581, Japan}

\begin{abstract}
We present a computational study of the compound 
Y(Co$_{1-x-y}$Fe$_x$Cu$_y$)$_5$ for  
0 $\leq x,y \leq 0.2$.
This compound was chosen as a prototype for investigating the cell boundary 
phase believed to play a key role in establishing the high coercivity
of commercial Sm-Co 2:17 magnets.
Using density-functional theory, we have calculated the magnetization and 
magnetocrystalline anisotropy at zero temperature for a range of compositions, 
modeling the doped compounds within the coherent potential approximation.
We have also performed finite temperature calculations for YCo$_5$, 
Y(Co$_{0.838}$Cu$_{0.162}$)$_5$  and Y(Co$_{0.838}$Fe$_{0.081}$Cu$_{0.081}$)$_5$
within the disordered local moment picture.
Our calculations find that substituting Co with small amounts of either Fe or 
Cu boosts the magnetocrystalline anisotropy $K$, but the change in $K$
depends strongly on the location of the dopants.
Furthermore, the calculations do not show a particularly large difference between the 
magnetic properties of Cu-rich Y(Co$_{0.838}$Cu$_{0.162}$)$_5$ and
equal Fe-Cu Y(Co$_{0.838}$Fe$_{0.081}$Cu$_{0.081}$)$_5$, despite
these two compositions showing different coercivity behavior when found
in the cell boundary phase of 2:17 magnets.
Our study lays the groundwork for
studying the rare earth
contribution to the anisotropy of Sm(Co$_{1-x-y}$Fe$_x$Cu$_y$)$_5$,
and also shows how a small amount of transition metal substitution 
can boost the anisotropy field of YCo$_5$.
\end{abstract}


\end{frontmatter}

\section{Introduction}
\label{sec.intro}

Of the wide variety of magnetic materials that can be formed
by alloying rare-earth elements with transition metals (RE/TM)~\cite{Buschow1977},
the permanent magnet market is dominated by
those based either on Nd-Fe-B~\cite{Sagawa1984,Croat1984} 
or Sm-Co~\cite{Strnat1967,Strnat1972}.
The performance of a permanent magnet is usually quantified
by its maximum energy product $(BH)_\mathrm{max}$, which measures
the energy stored in the air gap of the associated magnetic 
circuit~\cite{Chikazumi1}.
At temperatures up to approximately 120$^\circ$C, the Nd-Fe-B
materials have the highest $(BH)_\mathrm{max}$ of all available
permanent magnetic materials, but above this temperature 
their excellent performance sharply diminishes~\cite{Gutfleisch2011}.
By contrast, Sm-Co magnets do not have the same dramatic
sensitivity to temperature as Nd-Fe-B,
showing superior performance
above 120$^\circ$C~\cite{Gutfleisch2011} and
even operating at temperatures in excess of 
400$^\circ$C~\cite{Chen1998}.
Sm-Co magnets are therefore the materials of choice
for applications where high-temperature performance
is critical, e.g.\ sensing in manufacturing 
processes~\cite{Gutfleisch2009}.

The Sm-Co magnets can be further partitioned into
the 1:5 and 2:17 classes based on their nominal
crystal structures, with the highest-performing
magnets falling into the 2:17 class~\cite{Kumar1988}.
As well as Sm and Co, commercial 2:17 magnets also contain
Fe, Cu and Zr at 
an approximate stoichiometry Sm(Co$_{1-x-y-u}$Fe$_x$Cu$_y$Zr$_u$)$_z$,
where $z\sim7.5$, $x,y \sim 0.1$ and $u\sim0.01$~\cite{Liu1999}.
As illustrated by the value of $z$, the 2:17 magnets
also do not simply consist of a single Sm$_2$TM$_{17}$
phase but rather adopt a multi-phase structure~\cite{Rabenberg1982}.
This structure consists of a cellular phase composed of 
2:17 cells surrounded by thin ($\sim$10~nm) cell boundaries with an approximate 
1:5 stoichiometry, and a lamellar, Zr-rich ``Z'' phase~\cite{Hadjipanayis2000}.

It has long been accepted that this complex multi-phase
structure is essential to maintaining the excellent
high-temperature performance of the 2:17 magnets~\cite{Kumar1988}.
Recent work has highlighted the importance of the Z phase in aiding 
the formation of the cellular phase~\cite{Duerrschnabel2017}.
The critical role played by the cellular phase is then revealed
by electron microscopy experiments, which show the pinning
of magnetic domain walls at 2:17/1:5 boundaries~\cite{Fidler1982}.
This domain wall pinning inhibits magnetization reversal and
thus provides a coercive force.

Over the years a number of theories have been proposed
to explain the pinning of the domain walls~\cite{Fidler1982,
Kronmuller2002,Goll2000,Livingston1977,Xiong2004,Nagel1979,
Durst1988,Streibl2000,Yan2003,Gopalan2009,SepehriAmin2017}.
According to micromagnetic theory~\cite{Hubert}, the energy of a domain
wall depends both on the strength of the exchange
interaction $A$ and the magnetic anisotropy $K$ as $\propto \sqrt{AK}$.
Assuming that the 2:17 cells and 1:5 cell boundary phases
have different $A$ and $K$, there will be an energy barrier 
associated with a domain wall moving between 
these regions~\cite{Kronmuller2002}.
Interestingly, this argument does not rely on the 
domain wall energy being larger or smaller in the 
cell boundary phase compared to the cell~\cite{Goll2000}.
In models based on ``repulsive'' pinning, the domain
wall energy is higher in the cell boundary phase,
so the domain wall gets stuck in the cell~\cite{Livingston1977,
Kronmuller2002, Xiong2004},
while in ``attractive'' pinning models  the domain walls 
have higher energies in the cell, so they 
get pinned in the cell boundary phase 
instead~\cite{Nagel1979, Fidler1982, Durst1988,Streibl2000}.
More recently, models have been proposed where it is
the variation of $K$ within the cell boundary region
that determines the coercivity~\cite{Yan2003,Gopalan2009,
SepehriAmin2017}.

The existence of different models reflect the complicated
nature both of 2:17 magnets and of coercivity in general.
Indeed, the small size of the cell boundary phase
and of the domain walls themselves already presents a challenge
to continuum-based micromagnetics~\cite{Nagel1979}.
However, assuming micromagnetics can be
used to gain insight into the magnetization reversal process,
a basic question is what values 
of $A$, $K$ and magnetic polarization $J$ should be used
as input in the simulations.
Focusing on the magnetic anisotropy $K$, based upon
the values measured for SmCo$_5$ and Sm$_2$Co$_{17}$
one would expect a much larger anisotropy in the
1:5 cell boundary phase compared to the 2:17 cells.
Indeed, there are reports of measurements on a commercial 2:17
sample which support this view~\cite{Kronmuller2002}.
However, other experimental studies concluded that the 1:5 cell boundary was actually
softer (smaller anisotropy) than the cell~\cite{Durst1988}.
Another study found similar anisotropy energies for the two
phases, but explained the pinning of domain walls in terms
of a large difference in exchange energy $A$ between the phases~\cite{Nagel1979}.

Of course, a crucial property of the commercial magnets is
the presence of the additional elements Cu, Fe and Zr.
A recent 3D atom probe study measured the chemical 
compositions of the cell boundary phase for 2:17 magnets showing both high
and low coercivities, depending on heat treatment~\cite{SepehriAmin2017}.
This study reported that the high-coercivity sample coincided
with an enhanced Cu and diminished Fe content in the 1:5 cell boundary,
as well as a sharp interface between the cell and cell boundary phases.
Conversely, having a similar Fe and Cu content in the 1:5  cell boundary,
as well as having a diffusive interface between the cell and cell boundary phases, 
was correlated with low coercivity~\cite{SepehriAmin2017}.
The Zr content was found to be very small in both cases.

It would be desirable to be able to establish a link between
chemical composition and magnetic properties.
In Ref.~\cite{SepehriAmin2017} it is pointed out that, although
there is some data on binary Sm(Co,Cu)$_5$ (e.g.\ Ref.~\cite{Lectard1994}),
there is a gap in the literature considering the ternary
compound Sm(Co$_{1-x-y}$Fe$_x$Cu$_y$)$_5$.
Indeed we note that even the experimental data
of Ref.~\cite{Lectard1994} on Sm(Co$_{1-y}$Cu$_y$)$_5$
only reports anisotropy fields measured for $y\geq0.24$
($\geq$20\% by atom) which is already larger than the $\sim$15\% measured
in the cell boundary phase in Ref.~\cite{SepehriAmin2017}.
The modification of TM content by the substitution of Co with
Cu and Fe can be expected to change the magnetic anisotropy
of SmTM$_5$ in a number of ways, namely: (a) affecting the single
ion anisotropy of Sm by modifying the crystal field~\cite{Kuzmin2008}
(b) affecting the temperature dependence
of this single ion anisotropy by modifying the exchange
field felt by the RE due to the RE/TM interaction~\cite{Patrick20182}
and (c) modifying the contribution of the TM-3$d$ electrons
to the anisotropy\cite{Nordstrom1992,Daalderop1996,Steinbeck20012,Larson20042,Matsumoto2014}.
The introduction of Fe and Cu can also be expected
to affect the other key micromagnetic parameters 
$A$ and polarization $J$, leading for example to modified Curie
temperatures~\cite{Patrick2017}.
Furthermore, changing the local environment of the Sm
atoms may even modify their valency and orbital hybridization, further
affecting the anisotropy~\cite{Miyake2018}.

\begin{figure}
\centering
\includegraphics{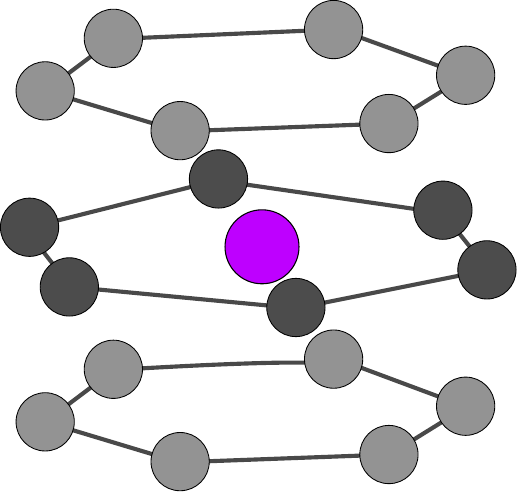}
\caption{Schematic representation of RETM$_5$
crystal structure (space group 191, $P6/mmm)$, showing the RE site (purple)
and two inequivalent TM sites: the $2c$ position (in plane with RE, dark grey),
and the $3g$ position (out of plane, light grey).
\label{fig.ballstick}}
\end{figure}
As a first step towards obtaining an improved understanding 
of how the magnetic properties of the cell boundary phase are
affected by chemical composition, we have performed first-principles
(density-functional theory) calculations on the ternary compound Y(Co$_{1-x-y}$Fe$_x$Cu$_y$)$_5$
(Fig.~\ref{fig.ballstick}).
The reasons for first investigating Y rather than Sm are twofold:
first, although Y has the same $s^2d$ valence structure as Sm,
the absence of 4$f$ electrons means that we can isolate the Tm-$3d$
contribution to the anisotropy [point (c) above] from the single ion contribution [(a) and (b)].
Second, YCo$_5$ remains an interesting magnet in its own right,
being free of lanthanide elements~\cite{Skokov2018}, having an anisotropy field of order 
20~T~\cite{Alameda1981} and potentially having a coercivity comparable to traditional SmCo$_5$
magnets~\cite{Gabay2014}.
Indeed, Fe-doped YCo$_5$ magnets are the subject of active research as potential
intermediate-performance permanent magnets~\cite{Soderznik2018}.

Our study consists of two parts.
In the first part, we calculate the zero-temperature
properties (magnetization and magnetocrystalline anisotropy)
of  Y(Co$_{1-x-y}$Fe$_x$Cu$_y$)$_5$ for 0 $\leq x,y \leq 0.2$.
The studied compositions fall into the ranges 
previously investigated in experiments on binary compounds~\cite{Wallace1979,TellezBlanco2000}.
The chemical disorder is modelled within the coherent
potential approximation (CPA)~\cite{Gyorffy1979}.
In the second part, we concentrate on the compounds
YCo$_5$, Y(Co$_{0.838}$Cu$_{0.162}$)$_5$ and
Y(Co$_{0.838}$Fe$_{0.081}$Cu$_{0.081}$)$_5$
and calculate their finite temperature properties
within the disordered local moment (DLM) picture~\cite{Gyorffy1985}.  
These particular concentrations were chosen based on the experimentally-measured
compositions of the cell boundary phases of (Sm-Co) 2:17 magnets
which showed high and low coercivity respectively in Ref.~\cite{SepehriAmin2017}.

The current manuscript aims to address the gap in the literature concerning
the intrinsic properties of the ternary  Y(Co$_{1-x-y}$Fe$_x$Cu$_y$)$_5$
compound.
From a technical aspect, due to the current lack of studies which have used 
the CPA to model the doping,
the manuscript includes some technical discussion, such as a comparison of the 
CPA with the simpler ``rigid band'' approach.
However, we also aim to make a practical connection to Ref.~\cite{SepehriAmin2017}
by reporting values of the anisotropy field and micromagnetic parameters
$A$, $K$ and $J$ for the representative high and low
coercivity cell boundary compositions.
We hope that such parameters might be useful for future micromagnetics
calculations like those originally performed in Ref.~\cite{SepehriAmin2017}.
In this way we follow recent works on RE/TM magnets which have 
demonstrated how microscopic quantities can be incorporated into large scale 
simulations~\cite{Toga2018,Fukazawa2019}.

Interestingly, the zero temperature calculations find 
that a low level of substitution enhances the magnetocrystalline
anisotropy, regardless of whether Co is substituted with Fe or Cu.
In particular, substituting $\sim$15\% of Co yields the largest
anisotropy energies.
The calculations also demonstrate how the anisotropy is very
sensitive to the location of the dopant atoms.

We find that the calculated difference
between Cu-rich and equal Cu-Fe substitution is not
particularly large.
The current calculations
therefore do not indicate that the TM-3$d$ contribution to the anisotropy of
the cell boundary phase is an important factor 
in determining the coercivity of the 2:17 magnets.
Nonetheless, our study lays the groundwork for studying the RE
contribution to the anisotropy of Sm(Co$_{1-x-y}$Fe$_x$Cu$_y$)$_5$,
and highlights the route of boosting the 
anisotropy field of YCo$_5$ through TM substitution.

Our manuscript is organized as follows.
In Section~\ref{sec.calc} we outline the methods used
to calculate magnetic properties at zero and finite
temperature.
The calculated results are presented in Section~\ref{sec.results}
and analyzed in Section~\ref{sec.discussion}.
We present our conclusions and discuss future directions
for study in Section~\ref{sec.conclusions}.

\section{Calculation details}
\label{sec.calc}

All calculations were performed within the multiple-scattering, 
Korringa-Kohn-Rostocker (KKR) formulation of density-functional theory (DFT)~\cite{Ebert2011},
treating exchange and correlation effects within the local
spin-density approximation (LSDA)~\cite{Vosko1980}.
Scalar-relativistic calculations were performed
within the atomic sphere approximation (ASA) for
the charge density and potential using the \texttt{Hutsepot}
KKR code~\cite{Daene2009}, solving the scattering
problem up to a maximum angular momentum quantum
number $l_\mathrm{max}=3$ and sampling the Brillouin
zone on a 20$\times$20$\times$20 grid.
The Y-4$p$ electrons were treated explicitly as valence
states.
The self-consistent potentials were obtained
for the T=0~K, ferromagnetic arrangement of
magnetic moments.

Substitutional doping of Co with Fe or Cu was modeled
within the CPA~\cite{Gyorffy1979,Ebert2011}.
For all compositions the lattice parameters were kept
fixed to the values $a$,$c$, = 4.950, 3.986~\AA,
as measured experimentally for YCo$_5$ at 300~K~\cite{AndreevHMM}.

To calculate magnetic properties,
the ``frozen'' scalar-relativistic
potentials were inserted into the Kohn-Sham-Dirac equation
in order to solve the fully-relativistic scattering problem~\cite{Strange1984}.
Spin and orbital magnetic moments were calculated by tracing
the appropriate operators with the Green's function~\cite{Strangebook}.
The magnetocrystalline anisotropy energy was obtained via
the torque, i.e.\ the change in free energy on rotation
of the magnetization vector~\cite{Wang1996,Staunton2006}.
An adaptive algorithm for the Brillouin zone integration
was used to ensure high numerical precision~\cite{Bruno1997}.

Finite temperature properties were calculated within
the disordered local moment picture, which treats the temperature-induced
fluctuations of the local moments at the level of the CPA~\cite{Gyorffy1985}.
The temperature-dependent Weiss fields were determined using an iterative 
procedure~\cite{Matsumoto2014,Patrick2017}.
Both an overview of DFT-DLM and the detailed procedure of evaluating the
Weiss fields and torque can be found elsewhere, e.g.\ Ref.~\cite{Staunton2006}.

Previous computational studies on YCo$_5$ found the LSDA to
yield values of both the orbital magnetic moments and the magnetocrystalline
anisotropy which are smaller than measured 
experimentally, but including an orbital polarization correction (OPC)~\cite{Eriksson19901} on the TM-$d$ orbitals
corrects this discrepancy~\cite{Nordstrom1992,Daalderop1996,Steinbeck20012}.
Although the relativistic DFT-DLM calculations do allow an OPC to 
be included~\cite{Patrick20182,Ebert1996,Patrick2018}, in the current work
we do not do so due to the large number of (Fe,Cu) compositions considered.
Therefore our calculated anisotropy energies are underestimates
compared to experiment.
Test calculations for selected compositions found the same qualitative 
trends to be obeyed by OPC and non-OPC calculations, but more work 
is necessary to perform a full comparison of the two approaches.

\section{Results}
\label{sec.results}

\subsection{Anisotropy at zero temperature: rigid band model}
\label{sec.rigidband}

For a crystal with hexagonal symmetry, the expected variation of the free energy  with magnetization
angle $\theta$ is  $K_1 \sin^2 \theta + K_2 \sin^4\theta + \mathcal{O}(\sin^6\theta)$,
where $\theta$ is given with respect to the $c$ axis.
Evaluating the torque $\partial F /\partial \theta$ at $\theta = 45^\circ$ yields $K_1 + K_2$,
which we label $K$.
Previous experimental and theoretical studies~\cite{Alameda1981,Patrick2018} 
have determined $K_2$ to be an order
of magnitude smaller than $K_1$ in pristine YCo$_5$, so $K \approx K_1$.
A positive value of $K$ corresponds to out-of-plane anisotropy.

The most straightforward method of simulating the
substitutional doping of Co with Fe or Cu is to use
the rigid band approximation, i.e.\ simply 
shift the Fermi level 
in the DFT calculation of
pristine YCo$_5$ 
so that the total number of 
electrons in the unit cell 
matches that expected for the doped system.
YCo$_4$Fe corresponds to a change
in electron number of $\Delta N_e = -1$, while
YCo$_4$Cu corresponds to $\Delta N_e = +2$.

\begin{figure}
\centering
\includegraphics{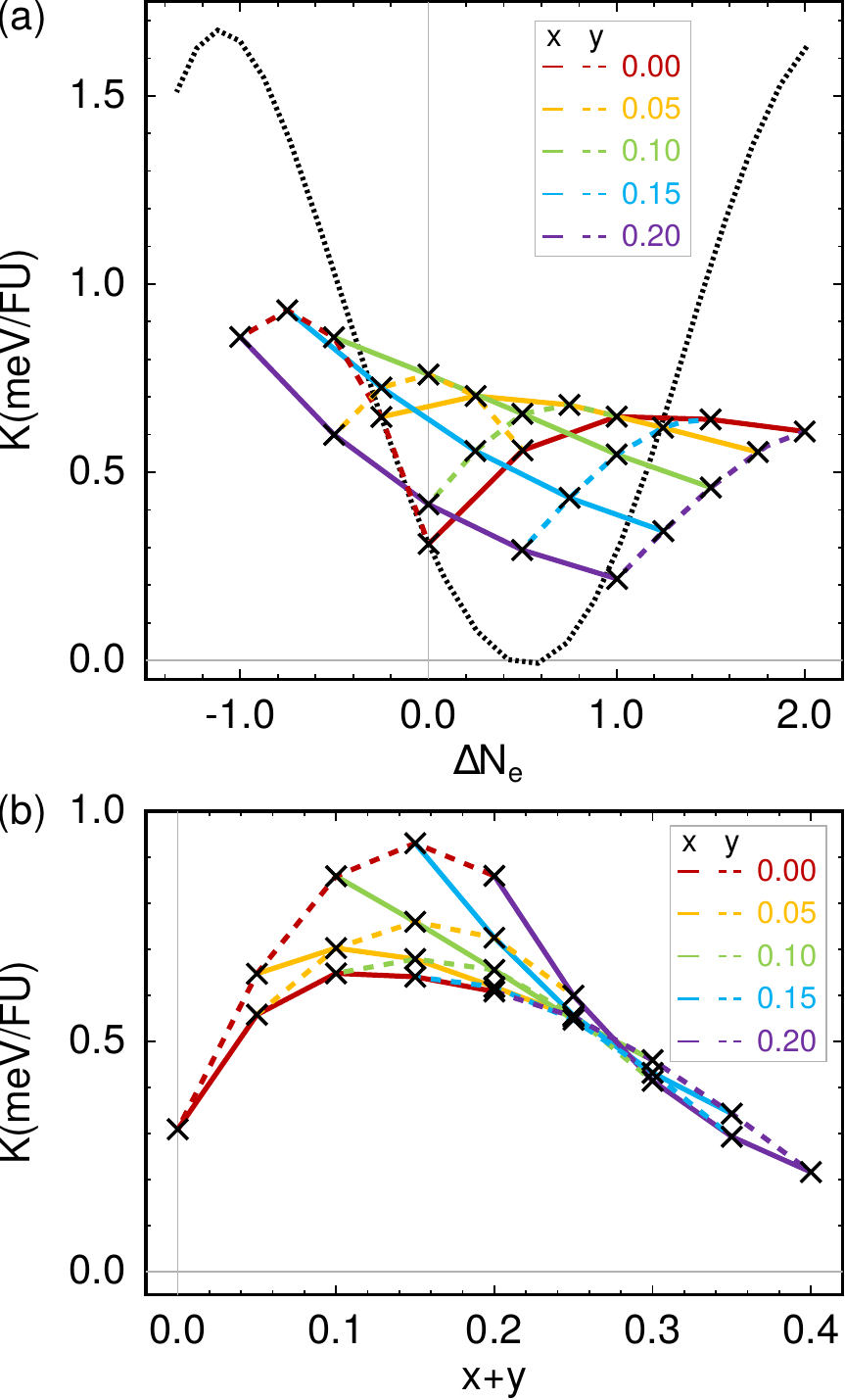}
\caption{
(a) Anisotropy energy $K$ per formula unit (FU)
as a function of change in electron number  $N_e$,
calculated for 
Y(Co$_{1-x-y}$Fe$_x$Cu$_y$)$_5$ either in the rigid band
approximation (black dotted line) or with the
CPA (crosses).  Each cross lies on the intersection of
a solid ($x$) and dashed ($y$) line which allows
the composition to be deduced.
(b) The same CPA calculations as (a) replotted as
a function of dopant content $x+y$.
\label{fig.KvN}}
\end{figure}
The anisotropy $K$ calculated in this way is shown as
the dotted line in Fig.~\ref{fig.KvN}(a).
As noted in previous works~\cite{Nordstrom1992,Daalderop1996,
Steinbeck20012,Matsumoto2014} there is a pronounced dependence
of $K$ on the band filling.
As discussed at length in Ref.~\cite{Daalderop1996}, the
anisotropy energy originates from the splitting of otherwise
degenerate states by the spin-orbit interaction, with the strongest
contributions coming from states with energies close
to the Fermi level.
Shifting the Fermi level changes the weights of the contribution
of each state, which may overall lead to an increase or decrease in $K$.

From the shape of the curve in Fig.~\ref{fig.KvN}(a) we see
that the rigid band model predicts that adding Fe would increase
$K$, up to a maximum close to YCo$_4$Fe.
By contrast, adding instead a small amount of Cu to form 
YCo$_{4.75}$Cu$_{0.25}$ would reduce $K$ to zero and yield
a perfectly soft magnet.
Increasing  the Cu content (e.g.\ YCo$_4$Cu) would again result in an
enhanced $K_1$ compared to the pristine case.

\subsection{Anisotropy at zero temperature: CPA, non-preferential substitution}

We now consider modeling the doping within the coherent potential 
approximation (CPA), a more sophisticated approach than the rigid
band model~\cite{Gyorffy1979}.
In these calculations it is necessary to specify the location
of the dopants, i.e.\ either the 2$c$ or 3$g$ crystallographic
sites (Fig.~\ref{fig.ballstick}).
Here we choose that Fe and Cu occupy 2$c$ and 3$g$ sites
with equal probability, i.e.\ non-preferential substitution, and 
explore the case that different sites
are preferred by different dopants in Section~\ref{sec.Kpref}.

We have calculated $K$ for 
Y(Co$_{1-x-y}$Fe$_x$Cu$_y$)$_5$ within the CPA
for $x,y = [0.00,0.05,0.10,0.15,0.20]$.
The data are shown in Fig.~\ref{fig.KvN} as crosses.
The composition of each data point can be deduced
by noting each cross lies at the intersection of
a solid and dashed line.
For instance,  the composition with the highest $K$,
Y(Co$_{0.85}$Fe$_{0.15}$)$_5$, lies at the intersection
of the $x=0.15$ (blue solid) and $y=0.00$ (red dashed)
lines.
The change in electron number $\Delta N_e$ is  -0.75.

Comparing the CPA and rigid band calculations shows
two key differences.
First, the predicted variation in $K$ is
smaller for the CPA case, with the CPA
values occupying a range of 0.8~meV/FU
compared to 1.7~meV/FU for the rigid band model.
Second, according to the CPA the addition of
dopants almost always increases $K$,
with only 
Y(Co$_{0.65}$Fe$_{0.20}$Cu$_{0.15}$)$_5$
and
Y(Co$_{0.60}$Fe$_{0.20}$Cu$_{0.20}$)$_5$
having a (slightly) reduced anisotropy energy
compared to the pristine case.
Therefore, the rigid band and CPA calculations
strongly disagree regarding the effect
of, for instance, adding a small amount of Cu.
Another example of the disagreement between
the two models is seen in configurations
with the same number of electrons, like
YCo$_5$, 
Y(Co$_{0.85}$Fe$_{0.10}$Cu$_{0.05}$)$_5$
and
Y(Co$_{0.70}$Fe$_{0.20}$Cu$_{0.10}$)$_5$.
According to the rigid band model, $K$ calculated
for each of these compounds should be the same,
but
in the CPA, $K$ varies over a range of 0.5~meV/FU.
In Section~\ref{sec.cparigid} we return to the comparison
of the rigid band model and CPA.

It is interesting to replot the CPA data
as a function of total dopant content, $x+y$,
which is done in Fig.~\ref{fig.KvN}(b).
In this case, the data for different ratios of 
Fe and Cu doping follow a more general trend,
which is an increase in $K$ up to a maximum
for $x+y$ = 0.15.
Plotting the data in this way indicates that 
the size of the Co deficit is an important
contributor to the variation in anisotropy.
Furthermore, replacing Co with Fe rather than Cu
yields the largest $K$ values, i.e.\ 
the compositions with
Cu content $y$ = 0.00.

\subsection{Magnetization and anisotropy field at zero temperature: 
CPA, non-preferential substitution}
\begin{figure}
\centering
\includegraphics{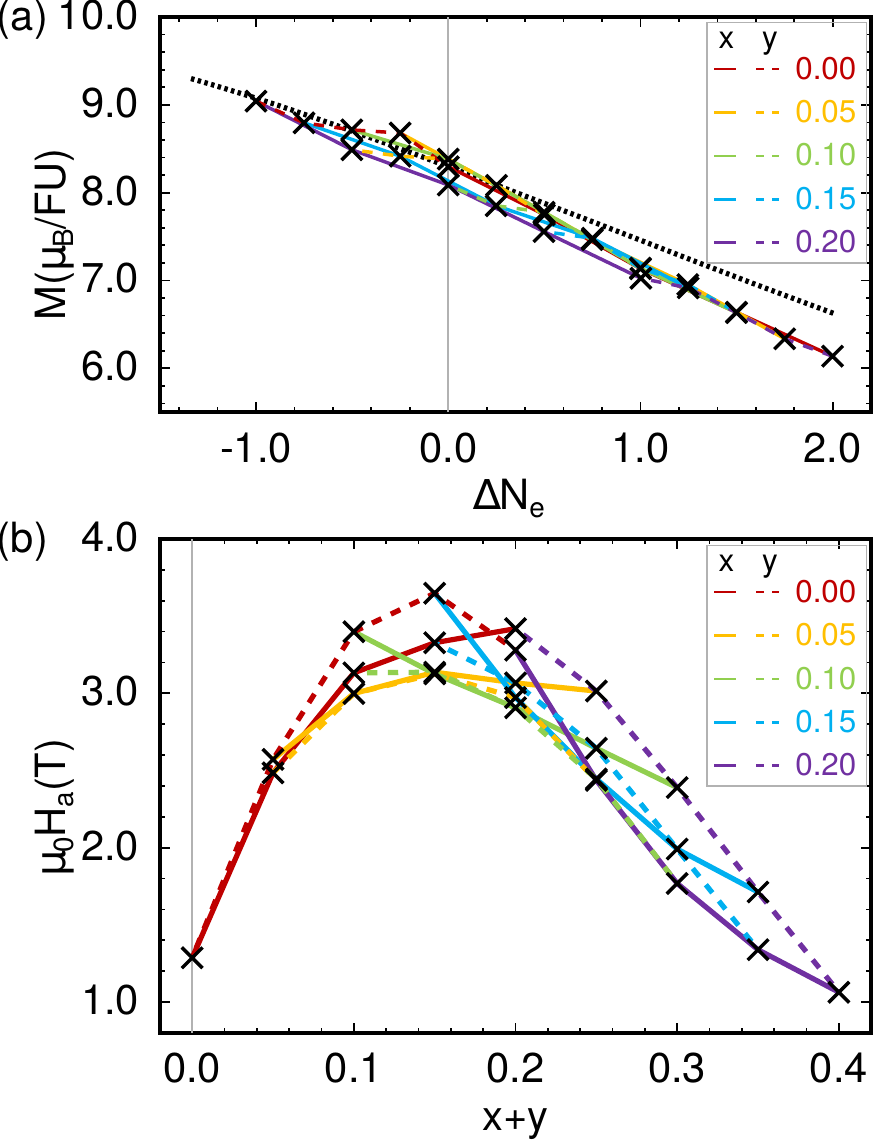}
\caption{ 
(a) Zero-temperature magnetization $M$ calculated as
a function of electron number within the rigid
band model (black dotted line) or using the CPA.
(b) Anisotropy field $\mu_0H_a$ ($= 2K/M$) obtained
in the CPA, plotted as a function of dopant
content $x+y$.
\label{fig.HavN}}
\end{figure}

According to micromagnetic theory, the theoretical
maximum for the coercive field of a ferromagnet is
the anisotropy field,  
$\mu_0H_a = 2K/M$~\cite{Brown1957,Kronmuller1987}.
Therefore, it is also important to investigate
the dependence of the magnetization $M$ on doping,
which is shown in Fig.~\ref{fig.HavN}(a).

In this case, there is quite close agreement
between the rigid band model (black dotted line)
and the CPA calculations.
The moment calculated for pristine YCo$_5$
is 8.3$\mu_B$/FU, where $\mu_B$ is the Bohr
magneton.
Adding electrons to YCo$_5$ (Cu-doping) increases the population
of the minority spin band, and therefore decreases
the magnetization.
The reverse applies when electrons are removed (Fe-doping).
We note that, although the agreement is generally good, 
the magnetization decreases
faster with Cu doping than predicted by the rigid band model.
For instance, for YCo$_4$Cu the magnetization calculated
with the CPA is 6.1~$\mu_B$/FU, compared to the rigid
band prediction of 6.6~$\mu_B$/FU.

Figure~\ref{fig.HavN}(b) shows the anisotropy field
calculated from the anisotropy energies and magnetizations
in Figs.~\ref{fig.KvN}(b) and \ref{fig.HavN}(a), respectively.
Comparing to  Fig.~\ref{fig.KvN}(b), we find a smaller scatter
in $H_a$ compared to $K$ for different compositions.
The reason for this smaller scatter is that the increased
$K$ from doping with larger amounts of Fe is partly offset
by the increased $M$; similarly, increasing the amount of
Cu weakens $M$ and therefore helps to boost $H_a$.
The clearest example is YCo$_4$Cu, whose value of $K$
is smaller than YCo$_4$Fe by a factor of 1.4 (0.61 vs.\ 0.86~meV/FU),
but whose anisotropy field is actually larger (3.4~vs.~3.3~T).
Nonetheless, the purely Fe-doped Y(Co$_{0.85}$Fe$_{0.15}$)$_5$ 
has both the highest anisotropy energy (0.93~meV/FU) 
and anisotropy field (3.6~T).

\subsection{Dependence of anisotropy on site occupation}

\begin{figure}
\centering
\includegraphics{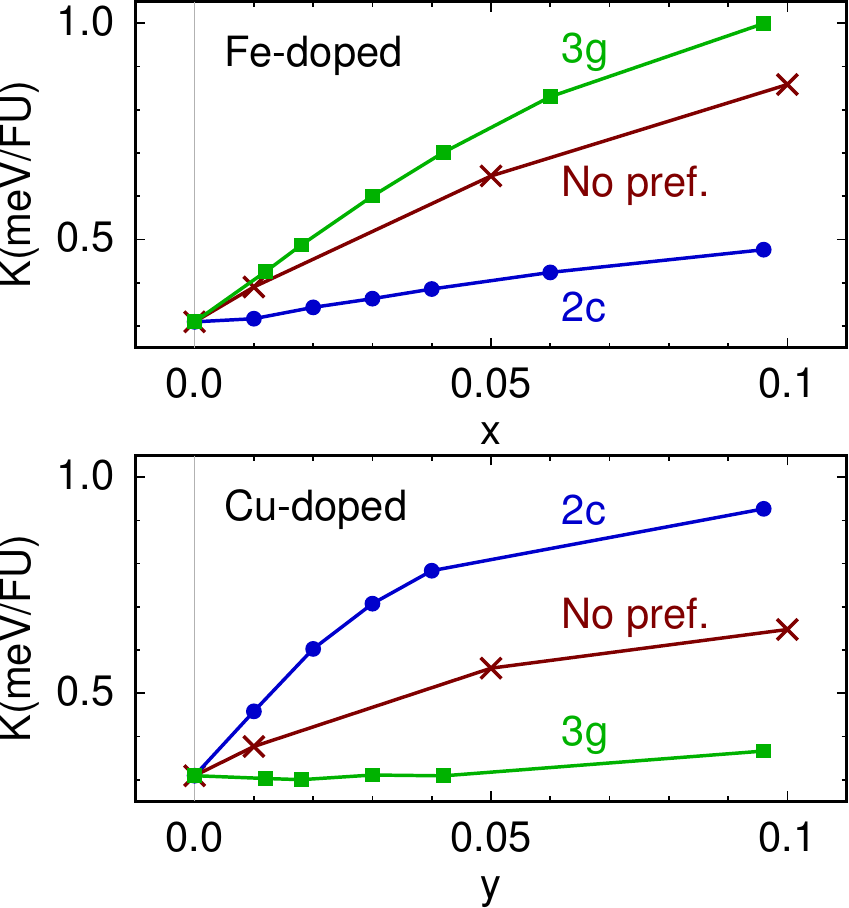}
\caption{Anisotropy energies of 
Y(Co$_{1-x}$Fe$_x$)$_5$ and
Y(Co$_{1-y}$Cu$_y$)$_5$ calculated with no preferential
(pref.) substitution at particular sites, compared
to exclusive substitution at the 2$c$ or 3$g$ crystal sites.
\label{fig.FeCu}}
\end{figure}

As stated already, the CPA calculations presented above
were performed assuming both Cu and Fe substitute
onto the 2$c$ and 3$g$ sites with equal probability.
On the other hand, it is possible that the dopants
may prefer to substitute at particular
sites, and that the value of $K$ might be affected.
Therefore, in Fig.~\ref{fig.FeCu} we compare the previous
calculations of the anisotropy energy of Y(Co$_{1-x}$Fe$_x$)$_5$ and
Y(Co$_{1-y}$Cu$_y$)$_5$
with the case where the dopants are placed exclusively
at the 2$c$ or 3$g$ crystal sites (Fig.~\ref{fig.ballstick}).

Interestingly, there is indeed a strong dependence of $K$
on the location of the dopants, with the site that gives 
the largest enhancement to the anisotropy depending on the dopant.
Substituting Co with Cu at the 3$g$ site has essentially
no effect on the anisotropy energy, while substituting 
with Fe at the 2$c$ site also does not result in a large
change.
By contrast, placing Cu at the 2$c$ site or Fe at
the $3g$ site has a large effect on $K$.
Substituting the dopants equally at both sites effectively
interpolates between the two limiting cases.

We are unaware of experimental data directly characterizing 
the location of dopant atoms in 
Y(Co$_{1-x}$Fe$_x$)$_5$ or Y(Co$_{1-y}$Cu$_y$)$_5$.
Neutron diffraction measurements on the related
compounds
Th(Co$_{1-x}$Fe$_x$)$_5$ and Y(Co$_{1-y}$Ni$_y$)$_5$
found preferential Fe/Ni substitution at the 3$g$/2$c$
sites, respectively~\cite{Deportes1976,Laforest1973}.
However based on the changes in lattice parameters
in Y(Co$_{1-x}$Fe$_x$)$_5$ with Fe doping it was
argued in Ref.~\cite{Rothwarf1973} that Fe preferred to occupy
2$c$ sites.
Our own CPA calculations performed at zero temperature (ferromagnetic
state) find $2c$ substitution to 
be preferred for Fe, Cu, and Ni~\cite{Patrick2017} but calculations
on stoichiometric systems including optimization
of the lattice parameters found Fe to prefer 3$g$
doping and Cu to prefer 2$c$~\cite{Uebayashi2002}.
Furthermore our test calculations  and previous work~\cite{Gabay2005}
have shown the energetics to be
sensitive to the modeling of the magnetic state (ferromagnetic versus
nonmagnetic).
For definiteness, 
 in what follows we place Fe
at 3$g$ sites and Cu at 2$c$ sites in line with the neutron
data~\cite{Deportes1976,Laforest1973}, the calculations including geometry optimization~\cite{Uebayashi2002}
and with previous theoretical works~\cite{Larson2004,Sakuma2006}.

\subsection{Anisotropy at zero temperature: CPA, preferential 3$g$/2$c$ substitution}
\label{sec.Kpref}
\begin{figure}
\centering
\includegraphics{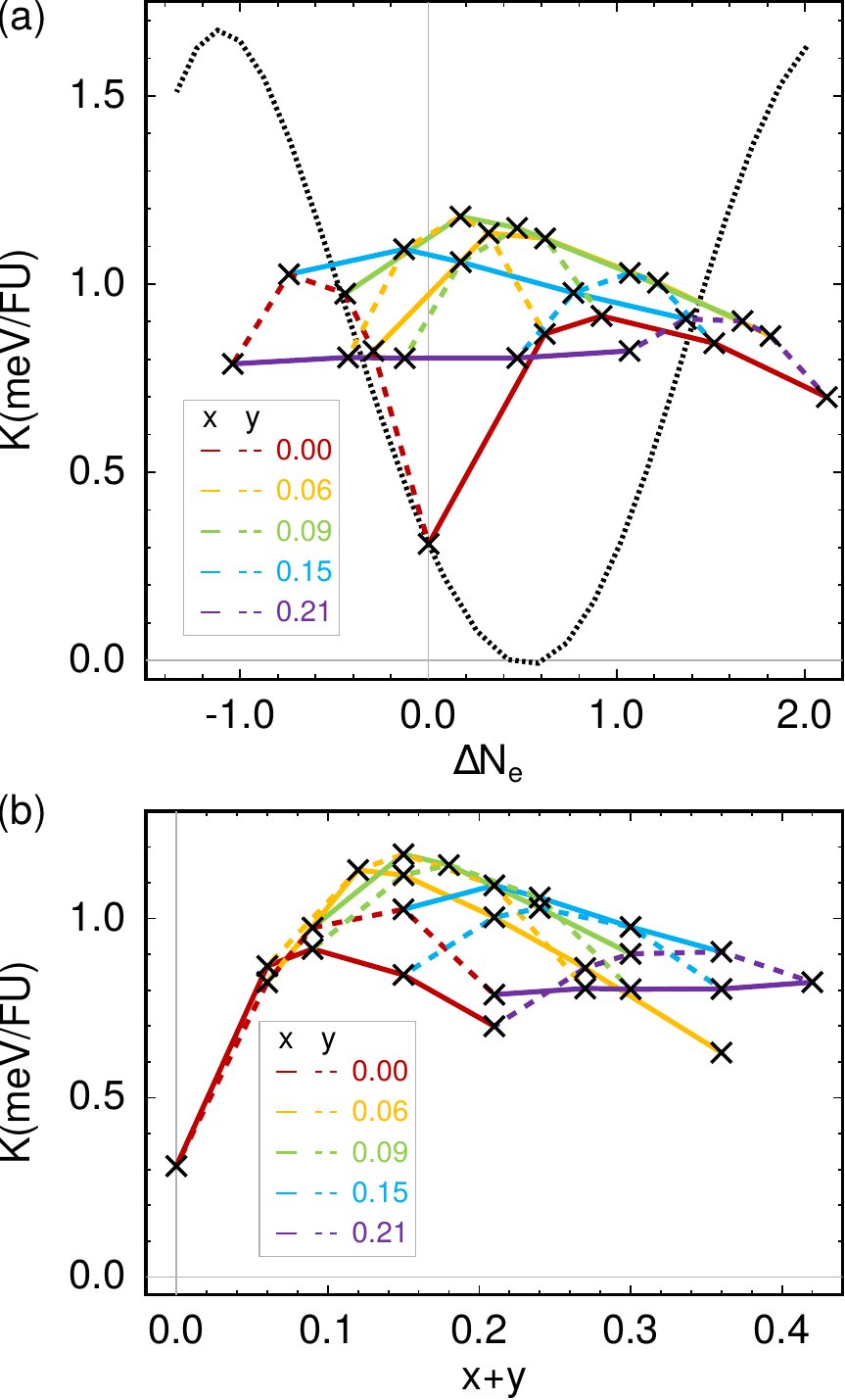}
\caption{ 
(a) Anisotropy energy $K$ of Y(Co$_{1-x-y}$Fe$_x$Cu$_y$)$_5$
calculated  either in the rigid band
approximation (black dotted line) or with the
CPA (crosses) (cf.\ Fig.~\ref{fig.KvN}).
The CPA calculations were performed placing Fe (Cu) at
3$g$ (2$c$) sites.
(b) The same CPA calculations as (a) replotted as
a function of dopant content $x+y$.
\label{fig.Kprefsub}}
\end{figure}

In Fig.~\ref{fig.Kprefsub}(a) we plot $K$ calculated
for Y(Co$_{1-x-y}$Fe$_x$Cu$_y$)$_5$ with the CPA, placing
Fe at the 3$g$ sites and Cu at the 2$c$ sites.
For comparison we show again the band-filling behavior
predicted by the rigid band approximation (Sec.~\ref{sec.rigidband}).

Referring to the data shown in Fig.~\ref{fig.KvN}(a), we see
that the anisotropy energy is enhanced compared to non-preferential
site substitution.
The largest value of $K$, calculated for
Y(Co$_{0.85}$Fe$_{0.09}$Cu$_{0.06}$)$_5$, is 1.18~meV/FU,
whilst the smallest value (apart from pristine YCo$_5$) is calculated
to be 0.70~meV/FU for
Y(Co$_{0.79}$Cu$_{0.21}$)$_5$.

As with the calculations with non-preferential site substitution,
the agreement between the rigid band and CPA calculations is
poor.
For instance, it is interesting to compare the sensitivity
of $K$ to the level of Cu doping, at low and high Fe content.
For Y(Co$_{1-y}$Cu$_{y}$)$_5$ (red solid line), $K$ varies
between 0.31~meV/FU at $y=0.00$ to a peak value
of 0.92~meV/FU at $y=0.09$.
However, including Fe at the level
Y(Co$_{0.79-y}$Fe$_{0.21}$Cu$_{y}$)$_5$
(purple solid line) reduces the variation in $K$
to just 0.03~meV/FU over the entire range of $y$.

Replotting the CPA calculations against $x+y$
[Fig.~\ref{fig.Kprefsub}(b)] does not show
as clear a trend as for the non-preferential
doping case [Fig.~\ref{fig.KvN}(b)].
However, again it is found that the largest
values of the anisotropy energy are calculated
for concentrations with $x+y\sim0.15$.
Around this optimal level, the largest values
of $K$ have similar Fe and Cu concentrations.
By contrast, the Cu-rich 
Y(Co$_{0.85}$Cu$_{0.15}$)$_5$ has a relatively low
$K$.
However, at the lowest dopant concentrations
$(x,y)=0.06$ the effect of replacing Co with either
Cu or Fe is very similar.

\subsection{Anisotropy fields at zero temperature: Cu-rich vs.\ equal Cu/Fe content}
\label{sec.Hapref}
\begin{figure}
\centering
\includegraphics{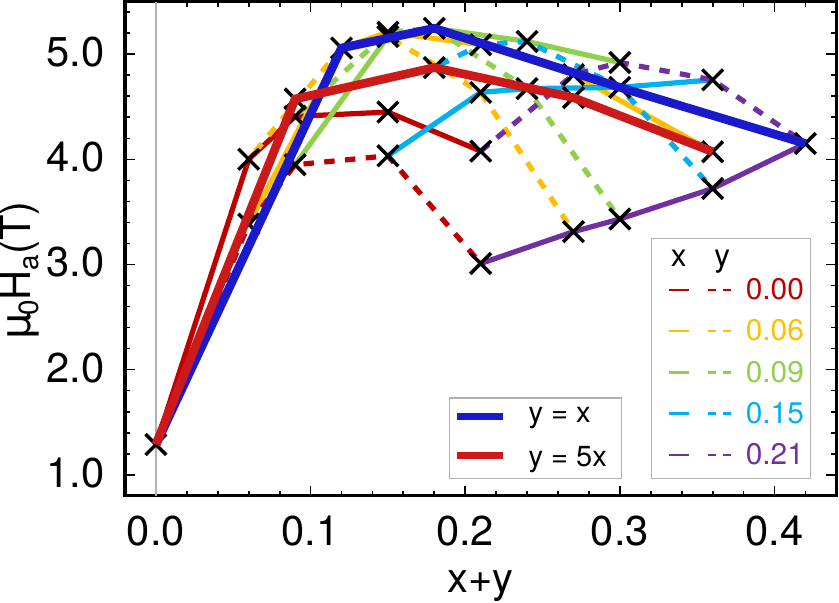}
\caption{
Anisotropy fields calculated within the CPA
placing Fe (Cu) at 3$g$ (2$c$) sites.
Additionally, calculations with equal Fe/Cu doping
($x=y$) and Cu-rich ($y=5x$) doping are shown with
thick blue and red lines, respectively.
\label{fig.Haprefsub}}
\end{figure}

Unlike the anisotropy energy, the magnetization $M$
is not particularly sensitive on the location of the
dopants, following the same behavior as shown in 
Fig.~\ref{fig.HavN}(a).
In Fig.~\ref{fig.Haprefsub} we plot the anisotropy
field for preferential substitution of
Fe/Cu at the $3g$/2$c$ sites.
As for the non-preferential case [Fig.~\ref{fig.HavN}(b)]
the anisotropy field is enhanced (weakened) for large Cu
(Fe) content, due to effect of the dopants on $M$.
This can be seen most clearly for the composition discussed
above,
Y(Co$_{0.79-y}$Fe$_{0.21}$Cu$_{y}$)$_5$,
which has $K$ effectively independent of $y$ but
$H_a$ which increases with Cu content
due to the corresponding reduction in $M$.

Motivated by the observations made at the end of the
previous section, 
Fig.~\ref{fig.Haprefsub}(b) shows some additional
data points, calculated for equal Fe/Cu doping
($x=y$; thick red line) and Cu-rich doping,
which we define as $y = 5x$  (thick blue line).
Here, it can be seen that despite the boost
to $H_a$ by the smaller magnetization of Cu, 
the compositions with equal amounts of Fe and Cu
have larger anisotropy fields than the Cu-rich
compositions.
Of all of the compositions considered,
Y(Co$_{0.82}$Fe$_{0.09}$Cu$_{0.09}$)$_5$
is found to have the highest anisotropy field
of $\mu_0H_a$ = 5.3~T.
The Cu-rich composition with the same $x+y$,
Y(Co$_{0.82}$Fe$_{0.03}$Cu$_{0.15}$)$_5$,
has $\mu_0H_a$ = 4.9~T.

\subsection{Anisotropy at finite temperature:
YCo$_5$, Y(Co$_{0.838}$Cu$_{0.162}$)$_5$ and
Y(Co$_{0.838}$Fe$_{0.081}$Cu$_{0.081}$)$_5$}

In order to make a tentative connection to the 2:17
Sm-Co magnets, we now focus on specific compositions
similar to those reported for the cell-boundary
phase in high and low coercivity samples in Ref.~\cite{SepehriAmin2017}
and calculate their properties between 0--300~K.
In Ref.~\cite{SepehriAmin2017}, high coercivity
was correlated with a Cu-rich cell-boundary phase,
while low coercivity was correlated with equal
Cu and Fe content.
We model the two cases with compositions  ($x$,$y$) = (0.00,0.162)
and (0.081,0.081).
We also compare to pristine YCo$_5$.
As in Secs.~\ref{sec.Kpref} and \ref{sec.Hapref} we place the
Fe and Cu dopants at the 3$g$ and 2$c$ sites, respectively.

\begin{figure}
\centering
\includegraphics{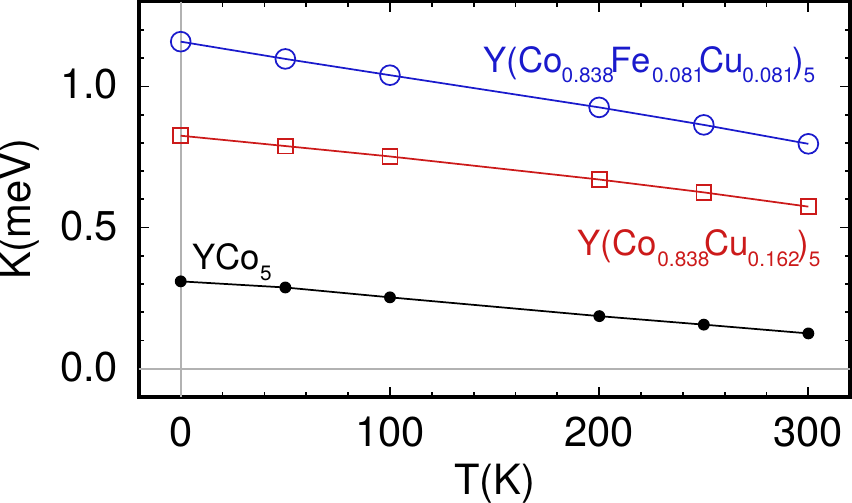}
\caption{Anisotropy energy $K$ calculated for
YCo$_5$ (black circles), Y(Co$_{0.838}$Cu$_{0.162}$)$_5$ (red squares) and
Y(Co$_{0.838}$Fe$_{0.081}$Cu$_{0.081}$)$_5$ (blue circles) as a function
of temperature.
\label{fig.KvT}}
\end{figure}

Figure~\ref{fig.KvT} shows the anisotropy energy $K$
calculated as a function of temperature $T$, for the
three cases.
As expected from Fig.~\ref{fig.Kprefsub}, at zero temperature
YCo$_5$ has the lowest $K$ value of 0.31~meV/FU, followed
by Y(Co$_{0.838}$Cu$_{0.162}$)$_5$ (0.83~meV/FU) and
then Y(Co$_{0.838}$Fe$_{0.081}$Cu$_{0.081}$)$_5$ (1.10~meV/FU).
Up to at least $T=300~K$, this ordering is unchanged,
with each composition showing a monotonic decrease in $K$
with temperature with a similar slope.
At 300~K, the $K$ values for the three configurations
are 0.12, 0.57 and 0.80~meV/FU respectively.

\begin{table}
%\scriptsize
\begin{tabular}{|ccccc|}\hline
($x,y$)      & $K$ (meV/FU) & $M$ ($\mu_B$/FU)  & $\mu_0H_a$ (T) &  $T_\mathrm{C}$ (K)\\\hline
(0.000,0.000) & 0.31 ; 0.12  & 8.14 ; 7.16      & 1.3 ; 0.6      & 841               \\
(0.000,0.162) & 0.83 ; 0.57  & 6.42 ; 5.52      & 4.4 ; 3.6      & 711               \\
(0.081,0.081) & 1.10 ; 0.80  & 7.63 ; 6.66      & 5.2 ; 4.1      & 788 \\
\hline
\end{tabular}
\caption{\label{tab.summary}
Anisotropy energy $K$, magnetization $M$, anisotropy field $H_a$ and Curie temperature $T_\mathrm{C}$ calculated
for 
Y(Co$_{1-x-y}$Fe$_{x}$Cu$_{y}$)$_5$.
The two values given for $K$, $M$ and $H_a$ correspond to calculations at $T$ = 0 and 300~K.
}
\end{table}

The magnetization and anisotropy field (not shown)
display the same monotonic decrease with temperature.
We also calculated the Curie temperatures $T_\mathrm{C}$, finding
the largest value (841~K) for pristine YCo$_5$.
The $T_\mathrm{C}$ of  
Y(Co$_{0.838}$Cu$_{0.162}$)$_5$ is found to be lower by over 100~K (711~K),
while  Y(Co$_{0.838}$Fe$_{0.081}$Cu$_{0.081}$)$_5$ lies in between (788~K).
The temperature-dependent results are summarized in Table~\ref{tab.summary}.

\subsection{Deriving parameters for micromagnetics simulations}

The quantities $K$, $M$ and $T_\mathrm{C}$ listed in Table~\ref{tab.summary}
may be obtained directly from the DFT-DLM calculations.
However micromagnetics simulations in fact require the magnetic polarization $J^\mathrm{m}$ 
and stiffness constant $A^\mathrm{m}$ in addition to the anisotropy $K^\mathrm{m}$~\cite{Hubert}.
Denoting the volume per formula unit as $\Omega (= \sqrt{3}a^2 c/2 )$, the anisotropy and
polarization can be straightforwardly related to the quantities given 
in Table~\ref{tab.summary}:
\begin{eqnarray}
K^\mathrm{m} &=&  K / \Omega \label{eq.K} \\
J^\mathrm{m} &=& \mu_0 M / \Omega 
\end{eqnarray}
We again note that the derived $K^\mathrm{m}$ values are likely to be underestimates
since no orbital polarization correction terms were included 
(Sec.~\ref{sec.calc}).

\begin{table}
%\scriptsize
\begin{tabular}{|cccc|}\hline
($x,y$)       & $K^\mathrm{m}$ (MJ/m$^3$) & $J^\mathrm{m}$ (T)  & $A^\mathrm{m}$ (pJ/m) \\\hline
(0.000,0.000) & 0.59 ;  0.23              & 1.12 ; 0.99         & 8.3 ; 6.6             \\
(0.000,0.162) & 1.57 ;  1.08              & 0.88 ; 0.76         & 7.0 ; 5.4             \\
(0.081,0.081) & 2.08 ;  1.52              & 1.05 ; 0.92         & 7.7 ; 6.1             \\
\hline
\end{tabular}
\caption{\label{tab.mm}
Anisotropy energy $K^\mathrm{m}$, magnetic polarization $J^\mathrm{m}$ and exchange 
stiffness constant $A^\mathrm{m}$ calculated
for Y(Co$_{1-x-y}$Fe$_{x}$Cu$_{y}$)$_5$ as discussed in the text.
The two values given correspond to calculations at $T$ = 0 and 300~K.
}
\end{table}
We have not yet established a formal framework to extract
the exchange stiffness constant $A^\mathrm{m}$ from the DFT-DLM calculations.
A similar challenge is encountered when performing calculations based on atomistic
spin models~\cite{Toga2018}.
In the current work we use a basic approximation~\cite{Hubert,Lectard1994}:
\begin{equation}
A^\mathrm{m} = (k_\mathrm{B}T_\mathrm{C} / \Omega) \ \zeta^2 \ [m_\mathrm{Co}(T)]^2
\label{eq.A}
\end{equation}
where $\zeta$ is the nearest neighbor distance between transition metal atoms
and $m_\mathrm{Co}(T)$ is the calculated order parameter of the Co moments
at temperature $T$.
The results of using equations~\ref{eq.K}--\ref{eq.A}
to express the quantities in Table~\ref{tab.summary}
as micromagnetics parameters are shown in Table~\ref{tab.mm}.

\subsection{Temperature in DFT-DLM calculations}
We conclude this section by noting that the classical statistical mechanics
used in the DLM picture leads to a faster decay of the the magnetic order
parameter $m_\mathrm{Co}(T)$ with temperature than observed experimentally~\cite{Patrick20183}.
As a result, the DFT-DLM ``temperature'' may in fact correspond
to a higher temperature in experiment.
To illustrate this aspect, we note from our DFT-DLM calculations on YCo$_5$ that at 
a calculated temperature of 300~K,  $m_\mathrm{Co}$ has a value of 0.90.
However, the experimental data shown in Ref.~\cite{Kuzmin2005} shows $m_\mathrm{Co} = 0.90$
in fact corresponds to a temperature of 465~K.
Conversely, at an experimental temperature of 300~K the order parameter is 0.95~\cite{Kuzmin2005},
which in the DFT-DLM calculations corresponds to a temperature of 200~K.
As illustrated here, it is relatively straightforward to correct for this
effect if required by mapping the DFT-DLM order parameters onto the experimental magnetization
versus temperature curves~\cite{Kuzmin2005}.

\section{Discussion}
\label{sec.discussion}

Here we discuss the specific results presented in Sec.~\ref{sec.results}
in more general terms.

\subsection{Modelling doping}
We first highlight some limitations of our chosen method
of simulating the doping.
First, we have not attempted to include structural effects
in our calculations, instead using the same lattice parameters
for all concentrations.
The anisotropy has previously been shown
to be sensitive to the $c$/$a$ ratio~\cite{Steinbeck2001}, so
different-sized dopants may affect $K$ indirectly in this way.
In the context of understanding the cell-boundary phase
in the 2:17 magnets modeling this effect is especially
difficult, since there may be local strains which invalidate
predictions based on Vegard's law or total energy minimization~\cite{Uebayashi2002}.

Also, the CPA assumes the limit of a dilute alloy, with
the dopants dispersed homogeneously throughout the host structure.
As a result, effects from short-range ordering are not included~\cite{Khan2016},
which may be especially important in understanding the low coercivity 2:17
Sm-Co sample which had a diffusive interface between the cell and cell boundary
phase~\cite{SepehriAmin2017}.

Finally, we note that we have assumed a perfect YCo$_5$ structure
and not explored the role of point defects, e.g.\ the ``dumbbell''
substitution which replaces Y with a pair of Co atoms~\cite{Kumar1988}.
It is possible that Fe and Cu may interact with these defects in different ways.

\subsection{Rigid band or CPA?}
\label{sec.cparigid}

Within the limitations of the above model, we explored two methods of modeling
doping, namely the rigid band model or the CPA.
While the two models produce similar values for the magnetization $M$, the predicted
values of $K$ differ substantially.
The rigid band model predicts much larger variations in the anisotropy
energy $K$ than the CPA.
Indeed the rigid band model predicts that the addition of a small amount of 
Cu should reduce $K$, potentially yielding a perfectly soft magnet
for  YCo$_{4.75}$Cu$_{0.25}$.  
By contrast the CPA almost always predicts $K$ to increase regardless of
the dopant species, especially if there is preferential substitution
at particular crystal sites.

From the theoretical point of view, the CPA is the more rigorous
approach~\cite{Gyorffy1979}.
As suggested by its name, the rigid band model cannot account
for changes in the bandstructure induced by the addition of dopants.
Furthermore if we take the view that the anisotropy depends not
only on the dopant species but also on which site it occupies 
(as argued experimentally a number of decades ago~\cite{Thuy1988}
and observed in the CPA calculations, e.g.\ Fig.~\ref{fig.FeCu}), we see that the
rigid band model cannot provide a full account of the behavior
of $K$.

\subsection{Comparison to experiment}

As noted in Sec.~\ref{sec.calc}, since the current calculations
do not include a correction for orbital polarization, we expect
the calculated values of $K$ to be smaller than observed
experimentally.
Therefore we restrict our comparison to trends in $K$ with doping.
We have not found experimental data on the ternary compound
Y(Co$_{1-x-y}$Fe$_x$Cu$_y$)$_5$, but studies on the binaries
Y(Co$_{1-x}$Fe$_x$)$_5$~\cite{Rothwarf1973,Wallace1979,Das2012} and
Y(Co$_{1-y}$Cu$_y$)$_5$~\cite{TellezBlanco2000} are available.

Considering Fe-doping first, Ref.~\cite{Rothwarf1973}
reports anisotropy constants at room temperature 
for $x$ = 0.0--0.3.
Here $K$ increases from a value of 4.5~MJm$^{-3}$ for $x$ = 0.0
to a maximum of 5.5~MJm$^{-3}$ at $x$ = 0.10, before reducing
again.
Interestingly a follow-up work~\cite{Wallace1979} states
that there were calibration errors the original data,
even though later papers continued to cite the original
reference~\cite{Thuy1988}.
A more recent work~\cite{Das2012} measured $K$ for
$x$ = 0.0--0.15 and found an increase from 4.2 to 5.0~MJm$^{-3}$,
(again at room temperature), reasonably consistent with
Ref.~\cite{Rothwarf1973}.

So, for Fe doping there is at least qualitative agreement between
different experiments and the current calculations
in predicting an increased $K$ on the addition of a small
amount of Fe.
Quantitatively,
the change of 0.8~MJm$^{-3}$ between $x=0.0$ and $x=0.15$ found
in Ref.~\cite{Das2012} corresponds to an absolute
increase of approximately 0.4~meV/FU, or a relative increase
of 19\%.
The (zero temperature) calculations predict an absolute
increase of 0.7~meV/FU over the same composition range 
(Fig.~\ref{fig.Kprefsub}), but a much larger relative increase
of over 300\%.

Now considering Cu doping, Ref.~\cite{TellezBlanco2000}
reports anisotropy fields for Y(Co$_{1-y}$Cu$_y$)$_5$
for $y$ = 0.0, 0.2 and 0.4, at temperatures $T > 200$~K
(lower temperature data are reported for $y=0.4$).
At the studied temperature range, YCo$_5$ is reported
to have the highest anisotropy field.
However the difference between the pristine case
and $y=0.2$ (YCo$_4$Cu) gets smaller with
decreasing temperature, and a straightforward linear 
extrapolation of the data indicates the anisotropy field of YCo$_4$Cu
would exceed YCo$_5$ at temperatures below 50~K.
We were unable to find experimental data measured for
lower Cu concentrations which could be compared more
directly to our calculations.
Such data, particularly at very low Cu concentration ($y\sim0.05$)
would be useful e.g.\ in comparing the rigid band model
with the CPA, since the former predicts a perfectly soft magnet
at this concentration (Fig.~\ref{fig.KvN}).

Finally, we note that although our calculations
should ideally be compared to anisotropy constants
measured for single crystals, there are a number
of experiments which report coercivity enhancement
in RE/TM magnets upon addition of Fe or Cu~\cite{Tozman2015,
Grechishkin2006,Buschow1976}.

\subsection{2:17 Sm-Co magnets}

We return to the original motivation of our work, to
quantify the effect on the TM contribution
to the anisotropy with Fe and Cu doping.
As shown by the blue and red lines in Figs.~\ref{fig.Haprefsub}
and~\ref{fig.KvT}, and the numbers reported in Table~\ref{tab.summary}
the differences between doping purely with Cu or with equal
quantities of Cu and Fe are not particularly large.
The anisotropy energy and field is calculated to be slightly
smaller for Y(Co$_{0.838}$Cu$_{0.162}$)$_5$  compared to
Y(Co$_{0.838}$Fe$_{0.081}$Cu$_{0.081}$)$_5$.
This fact, combined with a lower exchange stiffness constant for
Y(Co$_{0.838}$Cu$_{0.162}$)$_5$ (Table~\ref{tab.mm}) gives a lower 
domain wall energy in this phase compared to 
Y(Co$_{0.838}$Fe$_{0.081}$Cu$_{0.081}$)$_5$~\cite{Hubert}.

Applying this observation to a repulsive pinning model 
(where the 2:17 cells have smaller domain wall energies than the boundaries), 
one would expect stronger
pinning for boundaries with the 
Y(Co$_{0.838}$Fe$_{0.081}$Cu$_{0.081}$)$_5$ composition.
Unfortunately however, in Ref.~\cite{SepehriAmin2017} this composition
(equal Fe and Cu content) corresponded to the low, not high, coercivity
sample.
Therefore, according to our calculations the TM contribution to the anisotropy
does not account for the correlation of coercivity with the composition of the cell
boundaries reported in Ref.~\cite{SepehriAmin2017}.
Of course, as described in the Introduction the larger Sm contribution 
to the anisotropy may provide this missing link.
In the final section we outline future work aimed at exploring
the 2:17 magnets further.

\section{Conclusions and outlook}
\label{sec.conclusions}

The aim of this work has been to study how the magnetic anisotropy 
and magnetization of YCo$_5$ is affected when Co is substituted with
Fe and/or Cu.
We have demonstrated how two different approaches to modeling dopants ---
the rigid band model and the coherent potential approximation (CPA) --- give
rather different predictions.
We also showed how the results of the CPA calculations depend very strongly
on which crystal site the dopant atoms occupy.

Our CPA calculations found that the anisotropy of YCo$_5$ could be enhanced
by adding reasonably small amounts of Fe and/or Cu,
with the largest anisotropy field at zero temperature observed for the composition
Y(Co$_{0.81}$Fe$_{0.09}$Cu$_{0.09}$)$_5$.
Of the compositions studied, assuming preferential substitution of Fe
and Cu at 3$g$ and 2$c$ sites respectively, Fe-rich samples (i.e.\
Y(Co$_{0.79-y}$Fe$_{0.21}$Cu$_{y}$)$_5$) had the lowest anisotropy
fields, but these still exceeded the anisotropy field of YCo$_5$ at zero temperature.

On the theoretical side, the obvious next step is to study the RE contribution
to the anisotropy in Sm(Co$_{1-x-y}$Fe$_{x}$Cu$_{y}$)$_5$.
In this case it will be essential to properly account for the crystal
field effects in the calculations~\cite{Hummler1996} and analyse 
the effects of hybridization of the $4f$ states with their 
environment~\cite{Patrick20182,Miyake2018}.
It will be interesting to see whether addition of a small quantity of Fe or Cu
boosts the anisotropy like in YCo$_5$, or whether $K$ decreases
for all compositions.
In addition to the 1:5 calculations, a DFT-DLM characterization of the 
pristine bulk Y$_2$Co$_{17}$ and Sm$_2$Co$_{17}$ will be necessary
to build a full picture of the cellular phase.

On the experimental side, first considering YCo$_5$, in our view
the question of the site preference of the Fe-dopants has still 
not been conclusively answered.
Having knowledge of this aspect would be a useful test both
of using the CPA to calculate magnetic anisotropy, and also
of total energy calculations in general to predict the preferred
location of dopants.
Furthermore, additional data exploring the behavior of $K$ for
low Fe and Cu content, particularly for the ternary system,
would also be useful.
Complementary measurements on SmCo$_5$ are also required.
In particular, as pointed out in Sec.~\ref{sec.intro}, the experimental
data exploring Cu-doped SmCo$_5$ does not extend to the critical
$y<0.2$ region~\cite{Lectard1994}.
A full characterization of (Y,Sm)(Co$_{1-x-y}$Fe$_x$Cu$_y$)$_5$ 
for $(x,y) < 0.2$ would therefore be a valuable contribution to 
the permanent magnet literature.

\section*{Acknowledgments}
The present work forms part of the PRETAMAG project,
funded by the UK Engineering and Physical Sciences Research
Council, Grant No. EP/M028941/1.
MM’s work was partly supported by JSPS KAKENHI Grant Number 15K13525.
We thank H.\ Sepehri-Amin for useful discussions.

%\bibliography{papers}

\begin{thebibliography}{10}
\expandafter\ifx\csname url\endcsname\relax
  \def\url#1{\texttt{#1}}\fi
\expandafter\ifx\csname urlprefix\endcsname\relax\def\urlprefix{URL }\fi
\expandafter\ifx\csname href\endcsname\relax
  \def\href#1#2{#2} \def\path#1{#1}\fi

\bibitem{Buschow1977}
K.~H.~J. Buschow, Rep. Prog. Phys. 40 (1977) 1179.
\newblock \href {http://dx.doi.org/10.1088/0034-4885/40/10/002}
  {\path{doi:10.1088/0034-4885/40/10/002}}.

\bibitem{Sagawa1984}
M.~Sagawa, S.~Fujimura, N.~Togawa, H.~Yamamoto, Y.~Matsuura, J. Appl. Phys. 55
  (1984) 2083--2087.
\newblock \href {http://dx.doi.org/10.1063/1.333572}
  {\path{doi:10.1063/1.333572}}.

\bibitem{Croat1984}
J.~J. Croat, J.~F. Herbst, R.~W. Lee, F.~E. Pinkerton, J. Appl. Phys. 55 (1984)
  2078--2082.
\newblock \href {http://dx.doi.org/10.1063/1.333571}
  {\path{doi:10.1063/1.333571}}.

\bibitem{Strnat1967}
K.~Strnat, G.~Hoffer, J.~Olson, W.~Ostertag, J.~J. Becker, J. Appl. Phys. 38
  (1967) 1001--1002.
\newblock \href {http://dx.doi.org/10.1063/1.1709459}
  {\path{doi:10.1063/1.1709459}}.

\bibitem{Strnat1972}
K.~Strnat, IEEE Trans. Magn. 8 (1972) 511--516.
\newblock \href {http://dx.doi.org/10.1109/TMAG.1972.1067368}
  {\path{doi:10.1109/TMAG.1972.1067368}}.

\bibitem{Chikazumi1}
S.~Chikazumi, Physics of Ferromagnetism, 2nd Edition, Oxford University Press,
  1997.

\bibitem{Gutfleisch2011}
O.~Gutfleisch, M.~A. Willard, E.~Br\"uck, C.~H. Chen, S.~G. Sankar, J.~P. Liu,
  Adv. Mater. 23 (2011) 821--842.
\newblock \href {http://dx.doi.org/10.1002/adma.201002180}
  {\path{doi:10.1002/adma.201002180}}.

\bibitem{Chen1998}
C.~H. Chen, M.~S. Walmer, M.~H. Walmer, S.~Liu, E.~Kuhl, G.~Simon, J. Appl.
  Phys. 83~(11) (1998) 6706--6708.
\newblock \href {http://dx.doi.org/10.1063/1.367937}
  {\path{doi:10.1063/1.367937}}.

\bibitem{Gutfleisch2009}
O.~Gutfleisch, High-Temperature Samarium Cobalt Permanent Magnets, Springer US,
  Boston, MA, 2009, pp. 337--372.
\newblock \href {http://dx.doi.org/10.1007/978-0-387-85600-1_12}
  {\path{doi:10.1007/978-0-387-85600-1_12}}.

\bibitem{Kumar1988}
K.~Kumar, J. Appl. Phys. 63~(6) (1988) R13--R57.
\newblock \href {http://dx.doi.org/10.1063/1.341084}
  {\path{doi:10.1063/1.341084}}.

\bibitem{Liu1999}
J.~F. Liu, Y.~Zhang, D.~Dimitrov, G.~C. Hadjipanayis, J. Appl. Phys. 85 (1999)
  2800--2804.
\newblock \href {http://dx.doi.org/10.1063/1.369597}
  {\path{doi:10.1063/1.369597}}.

\bibitem{Rabenberg1982}
L.~Rabenberg, R.~K. Mishra, G.~Thomas, J. Appl. Phys. 53~(3) (1982) 2389--2391.
\newblock \href {http://dx.doi.org/10.1063/1.330867}
  {\path{doi:10.1063/1.330867}}.

\bibitem{Hadjipanayis2000}
G.~C. Hadjipanayis, W.~Tang, Y.~Zhang, S.~T. Chui, J.~F. Liu, C.~Chen,
  H.~Kronm\"uller, IEEE Trans. Magn. 36~(5) (2000) 3382--3387.
\newblock \href {http://dx.doi.org/10.1109/20.908808}
  {\path{doi:10.1109/20.908808}}.

\bibitem{Duerrschnabel2017}
M.~Duerrschnabel, M.~Yi, K.~Uestuener, M.~Liesegang, M.~Katter, H.-J. Kleebe,
  B.~Xu, O.~Gutfleisch, L.~Molina-Luna, Nat.\ Commun. 8~(1) (2017) 54.
\newblock \href {http://dx.doi.org/10.1038/s41467-017-00059-9}
  {\path{doi:10.1038/s41467-017-00059-9}}.

\bibitem{Fidler1982}
J.~Fidler, J. Magn. Magn. Mater. 30~(1) (1982) 58 -- 70.
\newblock \href {http://dx.doi.org/10.1016/0304-8853(82)90010-5}
  {\path{doi:10.1016/0304-8853(82)90010-5}}.

\bibitem{Kronmuller2002}
H.~Kronm\"uller, D.~Goll, Physica B 319~(1) (2002) 122 -- 126.
\newblock \href {http://dx.doi.org/10.1016/S0921-4526(02)01113-4}
  {\path{doi:10.1016/S0921-4526(02)01113-4}}.

\bibitem{Goll2000}
D.~Goll, H.~Kronm{\"u}ller, High-performance permanent magnets,
  Naturwissenschaften 87~(10) (2000) 423--438.
\newblock \href {http://dx.doi.org/10.1007/s001140050755}
  {\path{doi:10.1007/s001140050755}}.

\bibitem{Livingston1977}
J.~D. Livingston, D.~L. Martin, J. Appl. Phys. 48~(3) (1977) 1350--1354.
\newblock \href {http://dx.doi.org/10.1063/1.323729}
  {\path{doi:10.1063/1.323729}}.

\bibitem{Xiong2004}
X.~Xiong, T.~Ohkubo, T.~Koyama, K.~Ohashi, Y.~Tawara, K.~Hono, Acta Mater.
  52~(3) (2004) 737 -- 748.
\newblock \href {http://dx.doi.org/10.1016/j.actamat.2003.10.015}
  {\path{doi:10.1016/j.actamat.2003.10.015}}.

\bibitem{Nagel1979}
H.~Nagel, J. Appl. Phys. 50~(2) (1979) 1026--1030.
\newblock \href {http://dx.doi.org/10.1063/1.326100}
  {\path{doi:10.1063/1.326100}}.

\bibitem{Durst1988}
K.~Durst, H.~Kronm\"uller, W.~Ervens, Phys. Status Solidi A 108~(1) (1988)
  403--416.
\newblock \href {http://dx.doi.org/10.1002/pssa.2211080143}
  {\path{doi:10.1002/pssa.2211080143}}.

\bibitem{Streibl2000}
B.~Streibl, J.~Fidler, T.~Schrefl, J. Appl. Phys. 87~(9) (2000) 4765--4767.
\newblock \href {http://dx.doi.org/10.1063/1.373152}
  {\path{doi:10.1063/1.373152}}.

\bibitem{Yan2003}
A.~Yan, O.~Gutfleisch, T.~Gemming, K.-H. M\"uller, Appl. Phys. Lett. 83~(11)
  (2003) 2208--2210.
\newblock \href {http://dx.doi.org/10.1063/1.1611641}
  {\path{doi:10.1063/1.1611641}}.

\bibitem{Gopalan2009}
R.~Gopalan, K.~Hono, A.~Yan, O.~Gutfleisch, Scr. Mater. 60~(9) (2009) 764 --
  767.
\newblock \href {http://dx.doi.org/10.1016/j.scriptamat.2009.01.006}
  {\path{doi:10.1016/j.scriptamat.2009.01.006}}.

\bibitem{SepehriAmin2017}
H.~Sepehri-Amin, J.~Thielsch, J.~Fischbacher, T.~Ohkubo, T.~Schrefl,
  O.~Gutfleisch, K.~Hono, Acta Mater. 126 (2017) 1.
\newblock \href {http://dx.doi.org/10.1016/j.actamat.2016.12.050}
  {\path{doi:10.1016/j.actamat.2016.12.050}}.

\bibitem{Hubert}
A.~Hubert, R.~Sch\"afer, Magnetic Domains, Springer-Verlag, 1998.

\bibitem{Lectard1994}
E.~Lectard, C.~H. Allibert, R.~Ballou, J. Appl. Phys. 75~(10) (1994)
  6277--6279.
\newblock \href {http://dx.doi.org/10.1063/1.355423}
  {\path{doi:10.1063/1.355423}}.

\bibitem{Kuzmin2008}
M.~D. Kuz'min, A.~M. Tishin, Vol.~17, Elsevier B.V., 2008, Ch.~3, p. 149.

\bibitem{Patrick20182}
C.~E. Patrick, J.~B. Staunton, Phys. Rev. B 97 (2018) 224415.
\newblock \href {http://dx.doi.org/10.1103/PhysRevB.97.224415}
  {\path{doi:10.1103/PhysRevB.97.224415}}.

\bibitem{Nordstrom1992}
L.~Nordstrom, M.~S.~S. Brooks, B.~Johansson, J. Phys.: Condens. Matter 4~(12)
  (1992) 3261.
\newblock \href {http://dx.doi.org/10.1088/0953-8984/4/12/016}
  {\path{doi:10.1088/0953-8984/4/12/016}}.

\bibitem{Daalderop1996}
G.~H.~O. Daalderop, P.~J. Kelly, M.~F.~H. Schuurmans, Phys. Rev. B 53 (1996)
  14415--14433.
\newblock \href {http://dx.doi.org/10.1103/PhysRevB.53.14415}
  {\path{doi:10.1103/PhysRevB.53.14415}}.

\bibitem{Steinbeck20012}
L.~Steinbeck, M.~Richter, H.~Eschrig, Phys. Rev. B 63 (2001) 184431.
\newblock \href {http://dx.doi.org/10.1103/PhysRevB.63.184431}
  {\path{doi:10.1103/PhysRevB.63.184431}}.

\bibitem{Larson20042}
P.~Larson, I.~Mazin, J. Magn. Magn. Mater. 269~(2) (2004) 176 -- 183.
\newblock \href {http://dx.doi.org/10.1016/S0304-8853(03)00589-4}
  {\path{doi:10.1016/S0304-8853(03)00589-4}}.

\bibitem{Matsumoto2014}
M.~Matsumoto, R.~Banerjee, J.~B. Staunton, Phys. Rev. B 90 (2014) 054421.
\newblock \href {http://dx.doi.org/10.1103/PhysRevB.90.054421}
  {\path{doi:10.1103/PhysRevB.90.054421}}.

\bibitem{Patrick2017}
C.~E. Patrick, S.~Kumar, G.~Balakrishnan, R.~S. Edwards, M.~R. Lees,
  E.~Mendive-Tapia, L.~Petit, J.~B. Staunton, Phys. Rev. Materials 1 (2017)
  024411.
\newblock \href {http://dx.doi.org/10.1103/PhysRevMaterials.1.024411}
  {\path{doi:10.1103/PhysRevMaterials.1.024411}}.

\bibitem{Miyake2018}
T.~Miyake, H.~Akai, J. Phys. Soc. Jpn 87~(4) (2018) 041009.
\newblock \href {http://dx.doi.org/10.7566/JPSJ.87.041009}
  {\path{doi:10.7566/JPSJ.87.041009}}.

\bibitem{Skokov2018}
K.~Skokov, O.~Gutfleisch, Scr. Mater. 154 (2018) 289 -- 294.
\newblock \href {http://dx.doi.org/10.1016/j.scriptamat.2018.01.032}
  {\path{doi:10.1016/j.scriptamat.2018.01.032}}.

\bibitem{Alameda1981}
J.~M. Alameda, D.~Givord, R.~Lemaire, Q.~Lu, J. Appl. Phys. 52~(3) (1981)
  2079--2081.
\newblock \href {http://dx.doi.org/10.1063/1.329622}
  {\path{doi:10.1063/1.329622}}.

\bibitem{Gabay2014}
A.~Gabay, X.~Hu, G.~Hadjipanayis, J. Magn. Magn. Mater. 368 (2014) 75 -- 81.
\newblock \href {http://dx.doi.org/10.1016/j.jmmm.2014.05.014}
  {\path{doi:10.1016/j.jmmm.2014.05.014}}.

\bibitem{Soderznik2018}
M.~Soder\v{z}nik, M.~Korent, K.~\v{Z}agar Soder\v{z}nik, J.-M. Dubois,
  P.~Tozman, M.~Venkatesan, J.~Coey, S.~Kobe, J. Magn. Magn. Mater. 460 (2018)
  401 -- 408.
\newblock \href {http://dx.doi.org/10.1016/j.jmmm.2018.04.036}
  {\path{doi:10.1016/j.jmmm.2018.04.036}}.

\bibitem{Wallace1979}
W.~E. Wallace, E.~V. Ganapathy, R.~S. Craig, J. Appl. Phys. 50~(B3) (1979)
  2327--2329.
\newblock \href {http://dx.doi.org/10.1063/1.326990}
  {\path{doi:10.1063/1.326990}}.

\bibitem{TellezBlanco2000}
J.~C. Tellez-Blanco, R.~Grossinger, R.~S. Turtelli, E.~Estevez-Rams, IEEE
  Trans. Magn. 36 (2000) 3333--3335.
\newblock \href {http://dx.doi.org/10.1109/20.908790}
  {\path{doi:10.1109/20.908790}}.

\bibitem{Gyorffy1979}
B.~L. Gy\"orffy, G.~M. Stocks, Springer US, 1979, Ch.~4, pp. 89--192.

\bibitem{Gyorffy1985}
B.~L. Gy\"orffy, A.~J. Pindor, J.~Staunton, G.~M. Stocks, H.~Winter, J. Phys.
  F: Met. Phys. 15 (1985) 1337.
\newblock \href {http://dx.doi.org/10.1088/0305-4608/15/6/018}
  {\path{doi:10.1088/0305-4608/15/6/018}}.

\bibitem{Toga2018}
Y.~Toga, M.~Nishino, S.~Miyashita, T.~Miyake, A.~Sakuma, Phys. Rev. B 98 (2018)
  054418.
\newblock \href {http://dx.doi.org/10.1103/PhysRevB.98.054418}
  {\path{doi:10.1103/PhysRevB.98.054418}}.

\bibitem{Fukazawa2019}
T.~Fukazawa, H.~Akai, Y.~Harashima, T.~Miyake, J. Magn. Magn. Mater. 469 (2019)
  296 -- 301.
\newblock \href {http://dx.doi.org/10.1016/j.jmmm.2018.08.071}
  {\path{doi:10.1016/j.jmmm.2018.08.071}}.

\bibitem{Ebert2011}
H.~Ebert, D.~K\"odderitzsch, J.~Min\'ar, Rep. Prog. Phys. 74~(9) (2011) 096501.
\newblock \href {http://dx.doi.org/10.1088/0034-4885/74/9/096501}
  {\path{doi:10.1088/0034-4885/74/9/096501}}.

\bibitem{Vosko1980}
S.~H. Vosko, L.~Wilk, M.~Nusair, Can. J. Phys. 58 (1980) 1200--1211.
\newblock \href {http://dx.doi.org/10.1139/p80-159}
  {\path{doi:10.1139/p80-159}}.

\bibitem{Daene2009}
M.~D\"ane, M.~L\"uders, A.~Ernst, D.~K\"odderitzsch, W.~M. Temmerman,
  Z.~Szotek, W.~Hergert, J. Phys.: Condens. Matter 21 (2009) 045604.
\newblock \href {http://dx.doi.org/10.1088/0953-8984/21/4/045604}
  {\path{doi:10.1088/0953-8984/21/4/045604}}.

\bibitem{AndreevHMM}
A.~V. Andreev, Vol.~8, Elsevier North-Holland, New York, 1995, Ch.~2, p.~59.

\bibitem{Strange1984}
P.~Strange, J.~Staunton, B.~L. Gy\"orffy, J. Phys. C: Solid State Phys. 17~(19)
  (1984) 3355.
\newblock \href {http://dx.doi.org/10.1088/0022-3719/17/19/011}
  {\path{doi:10.1088/0022-3719/17/19/011}}.

\bibitem{Strangebook}
P.~Strange, Relativistic Quantum Mechanics, Cambridge University Press, 1998.

\bibitem{Wang1996}
X.~Wang, R.~Wu, D.-s. Wang, A.~J. Freeman, Phys. Rev. B 54 (1996) 61--64.
\newblock \href {http://dx.doi.org/10.1103/PhysRevB.54.61}
  {\path{doi:10.1103/PhysRevB.54.61}}.

\bibitem{Staunton2006}
J.~B. Staunton, L.~Szunyogh, A.~Buruzs, B.~L. Gyorffy, S.~Ostanin, L.~Udvardi,
  Phys. Rev. B 74 (2006) 144411.
\newblock \href {http://dx.doi.org/10.1103/PhysRevB.74.144411}
  {\path{doi:10.1103/PhysRevB.74.144411}}.

\bibitem{Bruno1997}
E.~Bruno, B.~Ginatempo, Phys. Rev. B 55 (1997) 12946--12955.
\newblock \href {http://dx.doi.org/10.1103/PhysRevB.55.12946}
  {\path{doi:10.1103/PhysRevB.55.12946}}.

\bibitem{Eriksson19901}
O.~Eriksson, B.~Johansson, R.~C. Albers, A.~M. Boring, M.~S.~S. Brooks, Phys.
  Rev. B 42 (1990) 2707--2710.
\newblock \href {http://dx.doi.org/10.1103/PhysRevB.42.2707}
  {\path{doi:10.1103/PhysRevB.42.2707}}.

\bibitem{Ebert1996}
H.~Ebert, M.~Battocletti, Solid State Commun. 98~(9) (1996) 785 -- 789.
\newblock \href {http://dx.doi.org/10.1016/0038-1098(96)00202-5}
  {\path{doi:10.1016/0038-1098(96)00202-5}}.

\bibitem{Patrick2018}
C.~E. Patrick, S.~Kumar, G.~Balakrishnan, R.~S. Edwards, M.~R. Lees, L.~Petit,
  J.~B. Staunton, Phys. Rev. Lett. 120 (2018) 097202.
\newblock \href {http://dx.doi.org/10.1103/PhysRevLett.120.097202}
  {\path{doi:10.1103/PhysRevLett.120.097202}}.

\bibitem{Brown1957}
W.~F. Brown, Phys. Rev. 105 (1957) 1479--1482.
\newblock \href {http://dx.doi.org/10.1103/PhysRev.105.1479}
  {\path{doi:10.1103/PhysRev.105.1479}}.

\bibitem{Kronmuller1987}
H.~Kronm\"uller, Phys. Status Solidi B 144 (1987) 385--396.
\newblock \href {http://dx.doi.org/10.1002/pssb.2221440134}
  {\path{doi:10.1002/pssb.2221440134}}.

\bibitem{Deportes1976}
J.~Deportes, D.~Givord, J.~Schweizer, F.~Tasset, IEEE Trans. Magn. 12~(6)
  (1976) 1000--1002.
\newblock \href {http://dx.doi.org/10.1109/TMAG.1976.1059185}
  {\path{doi:10.1109/TMAG.1976.1059185}}.

\bibitem{Laforest1973}
J.~Laforest, J.~Shah, IEEE Trans. Magn. 9~(3) (1973) 217--220.
\newblock \href {http://dx.doi.org/10.1109/TMAG.1973.1067699}
  {\path{doi:10.1109/TMAG.1973.1067699}}.

\bibitem{Rothwarf1973}
F.~Rothwarf, H.~A. Leupold, J.~Greedan, W.~E. Wallace, D.~K. Das, Int. J.
  Magnetism 4 (1973) 267--271.

\bibitem{Uebayashi2002}
K.~Uebayashi, K.~Terao, H.~Yamada, J Alloys Compd 346 (2002) 47 -- 49.
\newblock \href {http://dx.doi.org/10.1016/S0925-8388(02)00516-9}
  {\path{doi:10.1016/S0925-8388(02)00516-9}}.

\bibitem{Gabay2005}
A.~M. Gabay, P.~Larson, I.~I. Mazin, G.~C. Hadjipanayis, J. Phys. D: Appl.
  Phys. 38~(9) (2005) 1337.
\newblock \href {http://dx.doi.org/10.1088/0022-3727/38/9/002}
  {\path{doi:10.1088/0022-3727/38/9/002}}.

\bibitem{Larson2004}
P.~Larson, I.~I. Mazin, D.~A. Papaconstantopoulos, Phys. Rev. B 69 (2004)
  134408.
\newblock \href {http://dx.doi.org/10.1103/PhysRevB.69.134408}
  {\path{doi:10.1103/PhysRevB.69.134408}}.

\bibitem{Sakuma2006}
A.~Sakuma, J. Appl. Phys. 99~(8) (2006) 08J307.
\newblock \href {http://dx.doi.org/10.1063/1.2176316}
  {\path{doi:10.1063/1.2176316}}.

\bibitem{Patrick20183}
C.~E. Patrick, S.~Kumar, K.~G\"otze, M.~J. Pearce, J.~Singleton, G.~Rowlands,
  G.~Balakrishnan, , M.~R. Lees, P.~A. Goddard, J.~B. Staunton, J. Phys.:
  Condens. Matter 30 (2018) 32LT01.
\newblock \href {http://dx.doi.org/10.1088/1361-648X/aad029}
  {\path{doi:10.1088/1361-648X/aad029}}.

\bibitem{Kuzmin2005}
M.~D. Kuz'min, Phys. Rev. Lett. 94 (2005) 107204.
\newblock \href {http://dx.doi.org/10.1103/PhysRevLett.94.107204}
  {\path{doi:10.1103/PhysRevLett.94.107204}}.

\bibitem{Steinbeck2001}
L.~Steinbeck, M.~Richter, H.~Eschrig, J. Magn. Magn. Mater. 226–230, Part 1
  (2001) 1011 -- 1013.
\newblock \href {http://dx.doi.org/10.1016/S0304-8853(00)01189-6}
  {\path{doi:10.1016/S0304-8853(00)01189-6}}.

\bibitem{Khan2016}
S.~N. Khan, J.~B. Staunton, G.~M. Stocks, Phys. Rev. B 93 (2016) 054206.
\newblock \href {http://dx.doi.org/10.1103/PhysRevB.93.054206}
  {\path{doi:10.1103/PhysRevB.93.054206}}.

\bibitem{Thuy1988}
N.~Thuy, J.~Franse, N.~Hong, T.~Hien, J. Phys. Colloq. 49~(C8) (1988)
  C8--499--C8--504.
\newblock \href {http://dx.doi.org/10.1051/jphyscol:19888227}
  {\path{doi:10.1051/jphyscol:19888227}}.

\bibitem{Das2012}
B.~Das, B.~Balamurugan, W.~Zhang, S.~V. J.~S. R.~Skomski, E.S.~Krage,
  D.~Sellmyer, REPM'12 - Proceedings of the 22nd International Workshop on
  Rare-Earth Permanent Magnets and their Applications 3 (2012) 427.

\bibitem{Tozman2015}
P.~Tozman, M.~Venkatesan, G.~A. Zickler, J.~Fidler, J.~M.~D. Coey, Appl. Phys.
  Lett. 107~(3) (2015) 032405.
\newblock \href {http://dx.doi.org/10.1063/1.4927306}
  {\path{doi:10.1063/1.4927306}}.

\bibitem{Grechishkin2006}
R.~M. Grechishkin, M.~S. Kustov, O.~Cugat, J.~Delamare, G.~Poulin,
  D.~Mavrudieva, N.~M. Dempsey, Appl. Phys. Lett. 89~(12).
\newblock \href {http://dx.doi.org/10.1063/1.2347282}
  {\path{doi:10.1063/1.2347282}}.

\bibitem{Buschow1976}
K.~H.~J. Buschow, M.~Brouha, J. Appl. Phys. 47~(4) (1976) 1653--1656.
\newblock \href {http://dx.doi.org/10.1063/1.322787}
  {\path{doi:10.1063/1.322787}}.

\bibitem{Hummler1996}
K.~Hummler, M.~F\"ahnle, Phys. Rev. B 53 (1996) 3272--3289.
\newblock \href {http://dx.doi.org/10.1103/PhysRevB.53.3272}
  {\path{doi:10.1103/PhysRevB.53.3272}}.

\end{thebibliography}

\end{document}